\documentstyle[aps,multicol,psfig,epsf,epsfig]{revtex}

\begin{document}

\def\xv{{\bf x}}
\tightenlines

\title{Interface depinning versus absorbing-state phase transitions}
\author{Mikko Alava $^{1}$ and Miguel A. Mu\~noz$^{2}$}
\address{
$^1$ Helsinki University of Technology, 
Lab. of Physics, HUT-02105 Finland
\\
$^2$ 
Instituto de F{\'\i}sica Te\'orica y Computacional, Carlos I,
Universidad de Granada, Facultad de Ciencias, 18071-Granada, Spain.
}
\date{\today}

\maketitle

\begin{abstract}
According to recent numerical results from lattice models, 
the critical exponents of systems with many absorbing 
states and an order parameter coupled to a non-diffusive conserved field
coincide with those of the linear interface depinning model
within computational accuracy. 
In this paper the connection between absorbing state phase transitions and
interface pinning in quenched disordered media is investigated.
For that, we present a mapping of the interface dynamics in a disordered 
medium into a Langevin equation for the active-site density and show 
that a Reggeon-field-theory like description, coupled to an additional
non-diffusive conserved field, appears rather naturally. 
Reciprocally, we construct a mapping from a discrete model belonging 
in the absorbing state with-a-conserved-field class to a discrete 
interface equation, and show how a quenched disorder is originated.
 We discuss the character of the possible noise terms in both 
representations, and overview the critical exponent relations.
Evidence is provided that, at least for dimensions larger that
one, both universality classes are just two different representations
of the same underlying physics.
\end{abstract}
\pacs{PACS numbers: 05.10.Gg, 05.70.Ln, 68.35}

\begin{multicols}{2}
\narrowtext

\section{Introduction}
Phase transitions separating a non-trivial from a frozen 
phase, in which the dynamics is completely arrested,
appear in a large variety of situations in physics, as well as 
in many other disciplines \cite{Marro,Granada,Haye}.
 A central problem from a theoretical viewpoint is to
understand how the symmetries and conservation laws of the dynamics
are reflected in the categorization of models into
universality classes. 
There are two main general contexts in which this
type of frozen states appear:

 (i) Lattice models with discrete particles; typically particles originate
``activity'' and the frozen state, without activity is 
referred to as ``absorbing state'' \cite{Marro,Granada,Haye}.
This group appears in various
disguises as cellular automata \cite{DK},
reaction-diffusion systems \cite{Marro,Haye}, 
directed-percolation-type models \cite{Haye}, 
or the fixed energy ensemble of sandpile cellular automata \cite{FES}, 
among many other examples.

(ii) Elastic interfaces in random environments. 
In this second group, the dynamics is frozen whenever the 
interface is {\it pinned} by the disorder, while the non-trivial phase 
is the moving or depinned one \cite{HHZ,Barabasi}.

The number of physical realizations of both of these two generic 
families of phase transitions is huge 
\cite{Marro,Granada,Haye,HHZ,Barabasi}. 

 The most prototypical universality class in the first 
group is that embracing, among many other models and systems, 
directed percolation (DP) \cite{Marro,Granada,Haye,DK}.
At a continuous level the DP class 
is represented by the Reggeon Field Theory (RFT) \cite{RFT}, which can be
written in terms of the following  Langevin equation:
\begin{equation}
\partial_t \rho(\xv,t) =
a\rho - b\rho^2 + \nabla^2 \rho        
+ \sigma \sqrt{\rho} ~~ \eta(\xv,t)
\label{rft}
\end{equation}
where $\rho$ is an activity field, 
$a, b$, and $\sigma$ are constants and $\eta$ is a delta-correlated
Gaussian white noise. 
The RFT is the minimal field theory capturing the relevant ingredients
of the DP universality class. It can be renormalized using standard 
field theoretical methods and the associated critical exponents can be 
computed in $\epsilon$-expansion \cite{RFT}.
Other universality classes of absorbing-state phase transitions
have been identified; all of them owe their existence to the presence
of some additional symmetry or conservation law. Among them some 
example are: the conserved parity (CP) class, 
in which there are two $Z_2$-symmetric
equivalent absorbing states  \cite{BAW,Haye}, 
dynamical percolation \cite{dyp}, and
the different classes of transitions with extra 
conservation laws \cite{WOH,Dif,Romu1}.

In the group of pinned interfaces, the simplest continuous model 
for depinning is the quenched Edwards-Wilkinson (QEW) equation, 
also called, ``Linear Interface Model'' (LIM) 
\cite{HHZ,Barabasi} 
\begin{equation}
\partial_t h(\xv,t) = \nu \nabla^2 h (\xv,t) + F + \eta(\xv,h)~,
\label{lim}
\end{equation}
that describes an elastic interface (the Laplacian) at the
reference height $h(\xv,t)$, with surface tension $\nu$, 
under the influence of a constant external driving term 
$F$, and a {\em quenched noise} $\eta$. 
Equation (\ref{lim}) exhibits a {\em depinning transition} at a critical force
 $F_c$; the interface configuration and dynamics develop 
critical correlations in the vicinity of the critical point. 
The standard approach for a theoretical analysis of the LIM is 
the functional renormalization group method. One-loop expressions 
for the minimal set of exponents have been computed by 
Nattermann et al. \cite{Nattermann} on one hand, and by 
Narayan and Fisher \cite{NF} (see also the more
recent work by Le Doussal and collaborators \cite{Doussal}).
 Here one enters technically and conceptually difficult terrain
due to the renormalization of the whole disorder correlator.
The outcome is that for noise fields $\eta$ which do not
exhibit extra translational symmetries, the expected 
depinning behavior follows, very generally, that resulting from a 
random-field uncorrelated noise term: 
the LIM universality class \cite{Nattermann,NF,Lesch,LT}. 
Other universality classes in the interfaces-in-random-media realm 
are the quenched Kardar-Parisi-Zhang equation \cite{TKD,HHZ,Barabasi}
and the Edwards-Wilkinson equation with columnar noise
 \cite{columnar,Barabasi}.

Recent investigations (motivated by the analysis of sandpile 
models \cite{BTW,Manna}, the archetype of systems 
exhibiting self-organized criticality (SOC) \cite{SOC}) 
have demonstrated that different models showing a continuous transition 
into an absorbing phase and with an order parameter coupled linearly
to an extra, non-diffusive conserved field (NDCF) belong to a unique 
universality class \cite{Romu1,Romu2,AIP}, that we will refer 
to as NDCF class. 
This class differs from the extremely robust DP class owing to the 
presence of an additional conservation law \cite{Dif}. 
Moreover, the critical exponents of this class seem, 
within numerical accuracy, equal to those of the LIM class 
\cite{FES,Romu1,Romu2,AIP}.
This might be surprising at first sight, 
as in these models there is no quenched disorder, as there is in LIM, 
and disorder is usually a relevant perturbation when it
comes to universality issues.

From a different perspective this observation is not so surprising,
as different often tentative results have been reported in the literature
in order to relate the dynamics of sandpiles to that of elastic 
manifolds in random media \cite{previous,mapping}. 
Furthermore, there is one more viewpoint from which the coincidence 
between both types of models is not so striking, namely, that provided 
by the ``Run-Time Statistics'' technique \cite{marsili}. 
This technique or theory establishes that
quenched disorder can be mapped rather generically into long-range 
temporal correlations (i.e. a long-term memory) in the activity field, 
(note the idea works also the other way around) \cite{example}, and
has been recently applied with success to the Bak-Sneppen model among   
others \cite{BS}.
In the NDCF class the presence of 
a conserved field plays the role of a long-memory term, and therefore it
comes not as a big surprise that it is equivalent to a class with
quenched disorder.  

In this article we discuss in detail the relation between
the two presented groups of transitions, i.e. absorbing states with a 
conserved field and pinned interfaces in random media, including annealed 
(or thermal) and quenched disorder respectively. 
The connection between absorbing state models in the DP class 
(without a conserved field) and their interface representation 
has also been recently considered in the literature \cite{DM}. 
In particular the RFT was mapped into rather unusual interface equation,
not resembling any known interfacial problem.

The paper is structured as follows:
We start in Section \ref{ns} by presenting the RFT-like Langevin 
equation for the recently introduced NCDF class.
In \ref{fromLIM} we present a prototypical 
interface model in the LIM class, in particular the cellular automaton 
by Leschhorn \cite{Lesch} (see also \cite{LT})
and work out a derivation of a Langevin equation 
for the activity density $\rho$, paying particular attention 
to the way by which the noise can be found. 
In Section \ref{toLIM} we proceed conversely: we employ a discrete mapping 
of a model with absorbing states in the NDCF class into a continuous 
interface representation.
 We end up with an interface equation,
with several quenched noise terms that reflect the microscopic
rules and the thermal noise applied in them. We discuss at
this point the noise correlations that arise and their 
relevance, with the aid of the renormalization group (RG) literature.
Finally, we present a discussion and an appendix 
in which we outline the relations between the exponents in the
two different pictures.

\section{The NCDF field theory}
\label{ns}
One particular system in the NDCF class (out of the many studied 
\cite{Romu1,Romu2}) is a two-species reaction-diffusion model, 
in which one of the species in immobile \cite{WOH} (see section 
\ref{toLIM} for a detailed definition). 
It has the great advantage of allowing for a rigorous derivation 
of a coarse-grained field theory (or, equivalently, a Langevin equation)
via a Fock space representation of the dynamics
\cite{Fock,WOH,Romu2}.
The result is in the form of a Reggeon field theory coupled to an extra 
conserved non-diffusive field, or what is equivalent, a RFT
equation with an extra non-Markovian term \cite{Romu2,FES,AIP}. 
Quite remarkably this Langevin equation coincides (up to irrelevant terms)
with the one proposed previously, based only on symmetry and relevancy 
arguments, as the minimal Langevin equation capturing the physics of NDCF, 
namely \cite{FES,Romu2}:
\begin{eqnarray}
{\partial \rho(\xv,t) \over \partial t}  & = &
a \rho(\xv) - b \rho(\xv)^2 +
 \nabla^2 \rho(\xv,t) - \mu \psi(\xv,t) \rho(\xv,t) \nonumber \\
&&        
+ \sigma \sqrt{\rho(\xv,t)} \eta(\xv,t) \nonumber \\
{\partial \psi(\xv,t) \over \partial t}  & = &
   D  \nabla^2 \rho(\xv,t),
\label{ndcf}
\end{eqnarray}
plus higher order terms, irrelevant from naive power counting analysis
\cite{RG}.
Note that the second equation, describing the evolution
of the background conserved field (coarse grained representation of 
the total number of particles, which is conserved in the microscopic model),
represents an static non-diffusive field: in the absence of activity its
dynamics is frozen.
Observe also that the second equation, being linear, can be integrated out,
and a closed equation for the activity written down.
More concretely  
\begin{equation}
 \psi(\xv,t)= 
\psi(\xv,0) + D \int_0^t dt' \nabla^2 \rho(\xv,t').
\label{nonmarkovian}
\end{equation}
The first contribution in Eq(\ref{nonmarkovian}), 
a quenched (columnar) disorder, represents the initial condition, 
while the second is a non-Markovian term. 
 The Langevin equation (\ref{ndcf}), even though it looks rather
similar to the RFT, has resisted all renormalization attempts; 
therefore predictions about critical exponents coming from an 
epsilon expansion calculation are not available so far.

\section{Phenomenological activity description 
of LIM models}
\label{fromLIM}
We consider a representative of the LIM class, namely
the Leschhorn-Tang (LT) cellular automaton \cite{Lesch}. 
In order to study its relation with standard systems with absorbing states, 
we intend to cast it into a Langevin equation describing 
the evolution of an activity field \cite{Granada}.

The LT automaton is defined as follows. The interface field $h(\xv )$
satisfies at each discrete time step $t_i$ the following equation:
\begin{equation}
     h(\xv,t_{i+1}) = 
         \cases{ 
              h(\xv,t_i) + 1,  & $ f(\xv,t_i) > 0$ \cr
              h(\xv,t_i),      & $ f(\xv,t_i) \le 0$ \cr }
\end{equation}
where the force $f$ is given by the combination of elasticity and
a random quenched pinning force as
\begin{equation}
f(\xv,t_i) = \nabla^2 h (\xv,t_i) + \eta(\xv,h)
\end{equation}
where
$\nabla^2 h(\xv)$ is the discrete Laplacian, i.~e.~$\sum_{nn} 
h(nn) - 2D h(\xv)$ where $nn$ denotes the
nearest neighbors on a hyper-cubic lattice. A
reasonable choice for the noise is 
\begin{equation}
\label{limnoise}
  \eta(\xv,h) = \cases{
                     +1,  & $ p $ \cr
                     -1,      & $ 1-p$ \cr }
\end{equation}
when $p$ is a random number uniformly distributed between
zero and unity.
This choice implies that the average driving force
is $ F= \langle f \rangle  = 2p-1$. $F$ plays the
role of a control parameter. The critical point is  
estimated to be at $p_c \sim 0.800$ \cite{Lesch}.

 At every time step, and at each site where the
total driving force exceeds its threshold value, 
i.e., at each interface-site advance, we define
an activity variable and set it equal to one. 
On the other hand, in the remaining lattice 
sites the corresponding activity takes a zero value.       
Additionally, we also define
at each site and time, a continuous ``background'' variable,
equal to $\nabla^2 h(\xv,t)~+~F$.
This controls the probability of each interface site to advance
at each time, regardless of whether it actually slips or not.
Let us emphasize that this background variable is a 
conserved magnitude, i.e., it takes a constant value, equal to $F$
when integrated (summed) over the whole lattice. 
However, locally, it favors or inhibits the generation of new activity. 
We now build up a couple of equations for the evolution of the
two fields: the activity, $\rho(\xv,t)$, and the
background field, $\psi(\xv,t)$, which are the coarse grained field 
analogous of the previously defined site variables. 
Using the identification between activity and ready-to-advance
sites:  $h(\xv,t)=\int_0^t dt' \rho(\xv,t')~ + ~ h(\xv,0)$.
Let us write down a couple of {\it mean-field}  equations
for the two defined fields:
\begin{eqnarray}
{\partial \rho(\xv,t) \over \partial t} & =& 
- \rho(\xv,t) + 
 \rho(\xv,t) ~ {\cal{G}} ~[\psi(\xv,t)] 
\nabla^2 \rho(\xv,t) 
\label{rho} \\
\psi(\xv,t) & \equiv & \nabla^2 h(\xv,t) + F \nonumber\\
& =&\int_0^t dt \nabla^2 \rho(\xv,t) 
+  \nabla^2 h(\xv,0) + F
\label{psi}
\end{eqnarray}
The justification of the different terms is as follows:
\begin{itemize}
\item 
  The term $-\rho(\xv,t)$ describes the decay of active sites,
that after the corresponding interface-advance become,
 in general, non-active. At a coarse grained level higher order 
corrections, as $- b \rho^2(\xv,t)$ may also appear. 
In particular, they might play an important role 
in order to prevent the activity from growing unboundedly, i.e.
in stabilizing the theory.

\item 
  $+ \rho(\xv,t) {\cal{G}}[\psi(\xv,t)]$
represents the fact that activity is created in regions where
some activity is already present, and the rate of creation at each
point is a function of the local background field, $\psi(\xv,t)$.
 Observe that the total contribution of this term
when integrated over the whole space has to be zero, but locally it  
fosters or inhibits the creation of further activity. Again, higher
order powers of $\rho(\xv,t)$ might  also be included.

\item
  $ \nabla^2 \rho(\xv,t)$ describes the diffusion of 
activity. This terms appears generically
for diffusive systems at a coarse grained scale.

\item  In what respects the $\psi(\xv,t)$ field,  Eq.(\ref{psi}), 
we have just written its definition by equating $h(\xv,t)$
to the number of ``topplings'' (or activity events) 
at that point in all the preceding history, plus its
initial value.      

\end{itemize}

 Expanding ${\cal{G}}[\psi(\xv,t)]$ in power series, and keeping only
the leading contribution, we are left
with a term $+\lambda \rho(\xv,t)\psi(\xv,t)$
(where $\lambda$ is a constant) on the r.h.s. of Eq.(\ref{rho}) (observe that
the constant term in the Taylor expansion has to be zero as its integral 
has to be conserved, as argued before).
A posteriori, we shall show that the omitted terms, as well
as higher order corrections to the Laplacian term, are irrelevant in what
respects large scale, asymptotic, properties.
 
In order to account for the system fluctuations (completely ignored 
so far) we now introduce a noise field contribution to Eq.(\ref{rho}). 
For that, as it is well known in field theoretical descriptions of 
systems with absorbing states \cite{Granada,Haye}, a RFT noise term: 
$\sigma \sqrt{\rho(\xv,t)} \eta(\xv,t)$ is needed, where $\sigma$ is 
a constant and $\eta$ a Gaussian white noise.  
This just reflects the fact that, as $\rho$ is a local coarse grained 
variable its local fluctuations are proportional to its square-root 
(see \cite{Marro,Granada,Haye} and references therein).
It also captures the physical key ingredient: wherever activity 
vanishes locally, fluctuations are canceled \cite{Granada}.

  Before proceeding further, let us now discuss why the quenched disorder of
the microscopic model can be represented by an annealed noise in our
description. The key point is the observation that in active regions, i.e.
where the interface advances, a new noise variable is selected at every 
time step and, as the interface does not return to already 
passed regions, there is no need to store the microscopic noise history, 
and the noise can be freshly extracted from its probability distribution 
after every interface advance. 
Therefore, it is rather obvious that in depinned (active) regions,
 quenched and annealed noises are fully equivalent. 
More subtle is the connection of the two types of noises in
what respect pinned (absorbing) regions. While the annealed noise, $\eta$
changes in time even if there is no activity in a given region, 
its variations are completely irrelevant as the noise amplitude 
appears multiplied by $\sqrt{\rho}=0$. 
Noise (including its activity dependent amplitude) at a given spatial
 point changes only whenever activity arrives to it, mimicking perfectly  
what happens in the microscopic interface model, where pinned regions 
can be depinned only under the presence of neighboring moving regions.
Therefore, the considered time-dependent noise, reproduces properly
all the properties of the original quenched disorder.

All previous considerations lead finally to the following 
Langevin equation for the activity field:
\begin{eqnarray} && {\partial \rho(\xv,t) \over \partial t}  =
[-1+ \lambda F + \lambda \nabla^2h(\xv,0) ] \rho(\xv,t) 
+  \nabla^2 \rho(\xv,t)
 \nonumber \\
&& +  
\lambda \rho(\xv,t) \int_0^t dt' \nabla^2  \rho(\xv,t')  
+ \sigma \sqrt{\rho(\xv,t)} \eta(\xv,t)
\label{limLang}
\end{eqnarray}
where we have substituted $\psi$ by its expression coming
from Eq.(\ref{psi}).
In general, the system is expected to lose
memory of the initial state for long enough times, therefore
the dependence on   $\nabla^2 h(\xv,0)$ is expected to be washed out.
However, in some cases, as for instance one-dimensional systems,
due to the meager phase space, and the slow relaxation of the
initial condition, this might not be the case \cite{1d}. 

Performing a perturbative, diagrammatic study of the previous Langevin 
equation it is easy to see (already at one loop level) that a new
non-linearity (vertex), with the same degree of relevancy as the
nonlinear terms already present in the theory
(i.e. the non-local-in-time vertex and the noise one)
is perturbatively generated: $\rho^2(\xv,t)$. 
In fact, this term could have been introduced also at a mean field
level, as pointed out before, as a stabilizing term for the activity
equation.

 Including all the discussed terms into the equation for $\rho$, and 
 integrating the equation for $\psi$, we finally obtain:
\begin{eqnarray}
&&{\partial \rho(\xv,t) \over \partial t}  =
- a \rho(\xv,t) - b \rho(\xv,t)^2 
 +  \lambda
 \rho(\xv,t) \int_0^t dt' \rho(\xv,t') \nonumber \\
&&  + \lambda \nabla^2 h(\xv,0)+ \nabla^2 \rho(\xv,t)
+ \sigma \sqrt{\rho(\xv,t)} \eta(\xv,t)
\end{eqnarray}     
where $a = -1 +F \lambda $ and $b > 0$ are constants.
At this point, it is a rather straightforward exercise
to verify that no further relevant terms
are generated when including perturbative (diagrammatic)
corrections to the bare theory.
Therefore, {\it the resulting Langevin equation is identical to the one 
proposed for systems with an infinite number of absorbing states
and an activity field coupled to an static conserved field Eq.~(\ref{ndcf})}
\cite{FES,Romu1,Romu2}.

Summing up, we have mapped a 
microscopic model belonging in 
the LIM class to the Langevin equation characterizing the
NCDF class.
Though our derivation is not rigorous, we believe it provides
strong evidence that in fact LIM and NDCF define the same 
universality class.
 
\section{Mapping a reaction-diffusion model to depinning}
\label{toLIM}
In this section we proceed conversely to the previous one: 
starting from a microscopic model in the NDCF class
we map it onto the LIM continuous equation, Eq.(\ref{lim}).
To that end we follow a recipe already applied to many 
sandpile models exhibiting SOC \cite{mapping}.
Following \cite{Romu2} we consider a two-species reaction-diffusion 
process on a $L^d$ lattice, with particles of types  $A$ and $B$
involved.  At each site $i$, and at each (discrete) time step
 the following reactions take place:
\begin{eqnarray}
B_i        & \rightarrow & B_{nn}     \, ,\,\, r_d \equiv 1 \\
A_i + B_i & \rightarrow& 2 \times B_i \, ,\,\, r_1\\
B_i       & \rightarrow  & A_i        \, ,\,\, r_2 .
\end{eqnarray}
The $A_i$, $B_i$ denote particles of each kind at site $i$.
the $r$'s are the probabilities for the microscopic processes
to occur: diffusion, $r_d$, activation $r_1$, and passivation $r_2$.
 Without loss of generality we will
fix $r_d=1$, implying that, after having the chance to react,
$B$ particles diffuse with probability one.
Thus one can define a phase boundary between the active and 
absorbing phases in terms of the $r_1$, $r_2$ probabilities,
with a phase transition in-between. 
We assign occupation numbers $n_{A,i}$, $n_{B,i}$
to each site. As the $A$ particles are non-diffusive, this system has 
an infinite amount of absorbing states defined by $n_{B,i} =0$ for
all $i$, with $n_{A,i}$ arbitrary. 

Now we define (analogously to what is done for sandpiles 
\cite{mapping,FES}) a height field $H(\xv,t)$ which increases by one unit
every time a site gives one (or more than one) active, 
diffusing $B$ particle to one (or more than one) of its neighbors. 
When this happens we say, using the sandpile terminology,
that the site ``topples''.
In this way, the $H$-field measures the integrated activity at 
$\xv$ up to time $t$. 

 The mapping to an interface automaton with quenched noise
is based on the fact that both, the reactions between the $A$
and $B$ species, and the diffusion of particles can be accounted 
for by looking at their net effects at every time $B$ particles
leave the site $\xv$. 
One just have to look at $n_A$ and $n_B$ when the site becomes 
active and a particle diffuses out.
The dynamics of $H$ can be written as:
\begin{equation}
     H(\xv,t+1) = \cases{
                        H(\xv,t) + 1  & $ f(\xv,H) > 0 $ \cr
                        H(\xv,t),     & $ f(\xv,H) \le 0 $ \cr }
                   \label{evol}
\end{equation}
which is formally identical to the Leschhorn automaton in 
the LIM class, with  a local ``force'' defined as
\begin{equation}
f (\xv,H) = n_{tot}(\xv,H) - \xi (\xv,H)
\label{force}
\end{equation}
where  
$n_{tot}(x,H)= n_A (\xv,H) + n_B (\xv,H)$
is the total number of particles at $\xv$, and
$\xi$ is a local random threshold that results from the 
microscopic processes.
 More concretely, the noise $\xi$ is defined as follows:
Consider the site $\xv$ after the $H$-th toppling,
either $n_B (\xv,H) = 0$ or  $n_B (\xv,H) > 0$ (this last can be the case
if and only if particles have arrived from the nearest neighbors
at the same time step).
In the first case, it will remain zero until a particle arrives 
from a nearest neighbor site; then one is free to choose a
value for $\xi (\xv,H)$ such that it makes the force $f$ 
negative in the time interval between topplings $H$ and $H+1$.
In the second case, $n_B (\xv)$ will fluctuate owing to the microscopic 
passivation and activation processes, either going to 
$n_B(\xv)=0$ or inducing a toppling at the next time step.
The relative probabilities of these two alternatives, 
as derived from the microscopic dynamics,
are captured in the $\xi(\xv,H)$ probability distribution.

Observe that $\xi$ depends solely upon the total number of
particles after the preceding toppling and the 
microscopic dynamical rules. In particular, the larger $n_{total}$ 
the larger the probability to have many $B$ particles and the 
larger the probability to topple. 
Let us also remark that the immobile grains $n_A$ constitute 
a ``pinning force'' (the larger their relative number, 
the lesser the probability to topple). 
The point-wise noise field $\xi(\xv,H)$ should have weak two-point 
correlations in $\xv$
since, in particular, it is dependent on the number 
of grains received from the $nn$'s at the interface location $H(\xv)$ 
which induces weak site-site correlations. The fact that $n_A$ changes
slowly will make the $H$-part of the noise correlator 
$\langle \xi(\xv,H) \xi(\xv^{'} , H^{'} ) \rangle$
less trivial than a simple delta-function $\delta (H-H')$. 

Equation~(\ref{evol}) can be considered as a discrete
interface equation 
\begin{equation}
\frac{\Delta H}{\Delta t}=\theta \left(f(x,t) \right).
\label{eq:H-def}
\end{equation}
It can be rewritten with the help of two particle-fluxes:
$n_\xv^{in}$ and $n_\xv^{out}$, 
are the number of grains added to or removed from a given 
site $\xv$ up to time $t$, respectively. 
Let us also define $g$ as the average number of particles given
to the nearest neighbors at each toppling event. 
It is clear that for long enough times $n_{\xv}^{out} 
\approx g H(\xv)$; relative deviations 
from this equality being negligible asymptotically.
Defining also the average value of $n_{\xv}^{in}$, 
$ \bar{n}_\xv^{in}$, as
$\bar{n}_\xv^{in}= g / 2d\sum_{\xv_{nn}} H(\xv_{nn},t)$, 
we can compute a noise $\tau(\xv,H)$ 
as the deviation of $n_{\xv}^{in}$ with respect to its
 average value: 
\begin{equation}
\tau(\xv,t) =
 n_\xv^{in}- {g \over 2d} 
\sum_{\xv_{nn}} H(\xv_{nn},t).
\label{tau-const}
\end{equation} 
In other words, $\tau$ counts the relative proportion of 
particles diffused out from the neighbors that actually arrive
to the site under consideration, compared with its average value.
A site to which particles have toppled in excess will take a positive
value of $\tau$,  and therefore will be more likely to topple in 
the following time steps.  

Plugging this into Eq.(\ref{force}), and using that 
$n_{tot}(\xv,t) =n_{tot}(\xv,0) + n_\xv^{in} - n_\xv^{out}$,  
we can write \cite{mapping} 
\begin{equation}
f = {g \over 2d} ~ \nabla^2 H + F(\xv,0) - \xi(\xv,H) + \tau(\xv,H)
\label{force2}
\end{equation}
where $F(\xv,0) \equiv n_{tot}(\xv,t=0)$.

The discretization in Eq.(\ref{eq:H-def}) can be understood so that
the rules result in an {\it effective\/} force $f'$ that is exactly 
unity when the interface field $H$ advances. Thus
$\Delta H/\Delta t \equiv f' \, \theta(f) = f' \theta(f')$ \cite{mapping}.
This construction can be achieved by picking
$\xi$ to have exactly the right value in order to make the force
driving the interface equal to unity, if it is larger than zero.
One arrives finally at the discretized interface equation
\begin{equation}
\frac{\Delta H}{\Delta t}  = 
{g \over 2d} \nabla^2 H + F(\xv,0)- \xi(\xv,H) + \tau(\xv,H).
\label{efflim}
\end{equation}

Let us stress the presence of three different noise terms:

\begin{enumerate}
\item $F(\xv,0)$ represents the original total-number-of-particle
configuration at $t=0$, and is  therefore a {\it columnar noise} 
term \cite{columnar}. It induces an initial transient regime until 
eventually, the dynamics washes out the dependence of the original 
configuration. In general, columnar disorder is irrelevant
in the renormalization group sense as compared to quenched noise; therefore
using relevancy arguments, it could be eliminated, at least in 
high enough dimensions, close or above the critical one $d_c=4$. 
Notice that this statement is equivalent to the LIM symmetry,
by which static force fields $F(\xv,0)$ (independent of $H$)
is completely equivalent to the existence of a non-trivial
 initial interface profile $H(\xv, t=0)$.
However, in low dimensional
systems, and in particular in $d=1$, due to the meager phase space,
relaxation times might be huge, and the time needed to eliminate
the dependence on the initial particle distribution divergently large
\cite{1d}.

\item The noise term $\xi(\xv,H)$ represents the local threshold, 
determining whether a site with some B particles
topples at a given time or, alternatively, they are
transformed into A particles by microscopic processes. 
It captures the in-site microscopic dynamics, and depends essentially 
on $n_{tot}$, and on the microscopic probabilities. 
On a nutshell, it says how many of the $n_{tot}$ particles are of type $A$ 
after the microscopic dynamics has operated in the corresponding time step:
if all $n_{tot}$ are of type $A$ then $\xi > n_{tot}$, and $f <0$; 
conversely, if any of the particles is of type $B$ then $\xi < n_{tot}$ 
and $f > 0$.
Observe that if the diffusion probability was smaller than unity,
then we should substitute $\xi(\xv,H)$ by a ``thermal noise'' 
$\xi(\xv,t)$, i.e. $\xi$ would change its value after every
time step instead of changing only after each toppling:
this is due to the fact that if $r_d <1$ then a site $\xv$  
including B-type particles could not topple at time t ($\xi(\xv,t)$ 
below threshold), and do so at a future time $t'$ 
($\xi(\xv,t')$ above threshold). This ``thermal noise'' would generate
a rounding off of the transition, but the critical exponents 
should not be affected by this irrelevant perturbation \cite{Nattermann}. 
Therefore, we stick to the simplest case, $r_d=1$.

\item The noise term $\tau$ keeps track of the Brownian motion of 
particles; i.e. it takes into account the fact that particles are
not homogeneously distributed among the $nn$, but one of them is picked up
randomly for each toppling event. 
It changes slowly since the effect of the random choices (directions) 
on the configuration is slow. 
This is in particular true since the noise $\tau$ is conserving,
as the number of particles is conserved 
(and as can be seen by integrating Eq.~(\ref{tau-const})). 
A key point is that, analogously to what discussed 
in the preceeding section, the choice to give a particle to a certain 
neighbor can be taken to be ``quenched'', i. e. chosen in advance at $t=0$,
 or ``annealed'', i. e. decided on the spot.
The correlator of $\tau$ can be generically written as
\begin{equation}
\langle \tau(\xv^{'},H') \tau(\xv,H) \rangle
\sim f_\parallel (\xv^{'} - \xv) f_\perp (H'-H).
\label{taucorr}
\end{equation}
The (so far unknown) correlators $f_\parallel$ 
and $f_\perp$ reflect the discrete nature of the choices in the dynamics. 
Two microscopic reasons lead immediately to non-trivial
correlations in $\tau$: 

(a) The noises $\tau$ at the $nn$'s of site 
$\xv$ are correlated due to an exclusion effect:
If a site gives out a diffusing $B$ particle to a neighbor, 
then all the other neighbors are excluded. 
The actual coarse-grained noise correlations are harder 
to assess, since the fluctuations in the particle flux that 
$\tau$ measures make the interface to fluctuate, and thus a 
separable noise correlator as Eq.~(\ref{taucorr}) is hard to compute. 
The easiest way to analyze the correlations among the different 
sites is therefore to compute the noise correlator from numerics of the
microscopic model, using the noise definition Eq.~(\ref{tau-const}).
This programme has been pursued for sandpiles \cite{mapping}.

(b) At each site the noise follows the dynamics of a random walker. 
In fact, every time a nearest neighbor topples, the choice 
(give the particle to $\xv$ or to a different site) 
makes it so that $f_\perp \sim (H^{'} - H)^{1/2}$ since at every step
$\tau$ can go ``up'' or ``down'' with respect to the average. 
\end{enumerate}

Therefore, reciprocally to what done in the previous section,
we have mapped the reaction-diffusion process into 
an interface equation. The dynamics of this interface equation 
follows exactly the history of a reaction diffusion process, 
the details of which are mapped into the quenched noises $\xi$ and 
$\tau$, and a columnar noise $F(\xv,0)$. 
Let us remark that the existence of a conservation law 
has played a key role in order to obtain a Laplacian in 
Eq.(\ref{efflim}).

Finally, using standard renormalization group arguments about the relevancy of 
different operators, {\it we can eliminate higher order irrelevant
terms and noise-correlations, and then we are left with the LIM 
equation for point-disorder} \cite{Nattermann,Doussal} (see also
the Appendix).

 It must be emphasized, that the mapping works in both ways, 
it is evident that the noise construction can be inverted to 
yield a reaction diffusion process, that corresponds to an 
interface model, assuming that the original noise terms have 
the right correlation and conservation properties. 
The interface model Eq.(\ref{efflim}) resembles
very much the one that corresponds to the Manna sandpile
automaton, with the addition of the $\xi$-noise term which
is more point-disorder like than the $\tau$-term. 

  Summing up, reciprocally to what was done in the previous section, 
in this one, we have constructed a mapping between a microscopic
model in th NCDF class into the Langevin equation for the LIM class.

\section{Discussion}
\label{Discusion}
We have presented strong theoretical evidence that, rather generically, 
the universality class of systems with many absorbing states 
and order parameter coupled to a non-diffusive conserved field, 
the NDCF class, and that of the linear interface model with point-disorder 
coincide. 
This fact, already pointed out from numerical simulations 
\cite{Romu1,Romu2,FES} is true at least nearby the critical 
dimension $d_c=4$, where relevancy arguments are reliable. 
In low dimensional systems this equivalence could break down owing to the 
existence, for example, of slow decaying initial conditions \cite{1d}.
For the frozen configurations in the point-disorder LIM it is known that the
correlations of the forces $\eta(x,H)$ acting on the interface vanish. In the
case of NDCF models, like the Manna sandpile, such correlations (now
computed from the particle configuration in frozen configurations) 
may become non-zero: this is a future avenue for numerical studies, 
but hopefully this would be a irrelevant feature. 

Likewise, if one considers a noise field for the LIM (Eq.~(\ref{limnoise}))
with non-trivial (power-law) bare correlations in $x$ or $H$, it is
unclear at this point how these should be reflected in the construction
of a Langevin equation for the corresponding activity field, 
like Eq.~(\ref{limLang}). 
Correlations in the local forces (or ``activity thresholds'') will affect the
way the coarse-graining works. For instance, due to the noise
structure the pinned and still-active regions will be correlated.

In order to establish the connection between the two classes
we have mapped a discrete interface model into the Langevin 
equation characteristic of the NDCF class, and conversely mapped 
a discrete model in the NDCF into the well known Langevin equation 
describing the LIM class.  
In order to have a more rigorous proof, one should be able to
map one Langevin equation into the other, but this, being the 
Langevin equations coarse-grained representations of the microscopic
models, is not an easy task to fulfill,
and remains an open challenge.
 
Let us remark that a similar problem remains open;
namely, the rigorous connection between the Quenched KPZ 
\cite{TKD,HHZ} depinning transition and directed percolation 
depinning \cite{Barabasi,DPD} in two dimensional systems, 
(and to directed surfaces in higher dimensions \cite{BGM}).
It is clear from numerics, that indeed these two universality 
classes coincide, but a satisfactory proof is, to the best of our 
knowledge, still lacking. 

It was the hope, that the possibility of renormalizing the NDCF
Langevin equation using standard RG techniques, of problem from 
the RFT-like equation approach, could shed some 
light on the (in principle, technically more difficult and obscure) 
functional renormalization group analysis of the interface equation with 
quenched noise.  
However, the difficulties encountered in renormalizing,
using standard perturbative schemes, the
Langevin equation for NDCF \cite{Romu2,AIP} are considerable; 
and have made all the attempts to renormalize
the theory to fall through.
Renormalizing the NDCF Langevin equation and relating the
derived critical exponents to those obtained using functional RG for
LIM remains an open and very challenging problem.

Finally let us also point out that all the discussions presented in 
this work deal with the ``constant force'' (in the interface language)
or ``fixed energy'' (in the absorbing-state terminology) ensemble.
They can be easily extended to the ``constant force'' or ``slow driving''
ensemble \cite{FES,mapping}, in which the system self-organizes 
into its critical state. This point is, however, not essential
since all evidence points to the fact that if two models belong to the
the universality class, they continue to share the same set of
critical exponents upon changing ensemble.

\vspace{1cm}
{\bf \centerline{APPENDIX}}
\vspace{0.5cm}

The scaling of the phase transition in the absorbing-state
representation is characterized by the exponents $\nu_\perp$, 
$\nu_\parallel$, $z$ and $\beta$. 
These describe the correlations in the activity
$\rho$ in the spatial and time directions, the development of
the correlations in time, and the behavior of $\rho$ above
the critical point, respectively. 
One has the scaling relation
\begin{equation}
\overline{\rho} (\Delta,L) = 
L^{-\beta/\nu_{\perp}} {\cal R} 
(L^{1/\nu_{\perp}} \Delta) \;,
\label{actfss}
\end{equation}
\noindent where $\Delta$ is the distance to the critical
point, and ${\cal R}$ is a scaling function with
${\cal R}(x) \sim x^{\beta}$ for large $\xv$.
For $L \gg \xi\sim \Delta^{-\nu_{\perp}}$ we expect
$\overline{\rho_a}\sim \Delta^\beta$ 
(here $\xi$ is the correlation length).  
When $\Delta=0$ we have that $\overline{\rho_a} (0,L) 
\sim L^{-\beta/\nu_{\perp}}$. For $\Delta>0$, by contrast,
$\overline{\rho_a}$ approaches a stationary value, while for 
$\Delta<0$ it falls off as $L^{-d}$. These can be used to
establish the numerical values of the exponents.

In the interface representation the relevant exponents are
$\nu$, $z$ as above, with the convention that $\nu \equiv \nu_\parallel$.
Usually it is assumed that the dynamics is self-affine, which implies
that $\nu_\perp = \chi \nu_\parallel$ \cite{HHZ,Barabasi}.
 This defines the roughness exponent $\chi$ that characterizes 
the spatial correlations of the interface. 
If ``simple scaling'' \cite{FV,Barabasi} holds, then one has a unique
roughness exponent and we can write for the interface width $w$
\begin{equation}
W^2 (t,L)  \sim \cases{
t^{2 \beta_W} \;, & $ t \ll t_{\times} $ \cr
L^{2 \alpha}\;, & $ t \gg t_{\times}, $  \cr}
\label{w2scal}
\end{equation}
using also the early-time exponent $\beta_w$.  
If simple scaling holds, we have the exponent relation 
$\beta_w z = \alpha$ \cite{FV}. If only one timescale is present,
the growth exponent  is related to the activity time-decay
exponent, $\theta$, via $\theta + \beta_W = 1$ \cite{DM}.

For point-like disorder the first-loop functional renormalization
group result reads $\chi = (4-d)/3$, and $z = 2 - (4-d)/9$
\cite{Nattermann}; see the extension to second order in \cite{Doussal}. From 
these, using the exponent relations, the other exponents follow. 
For rather generic bare disorder correlators the implication
is that the full correlator flows in the renormalization to this
``random field'' (or point-disorder)  fixed point function, 
and thus the exponents are the same.  
However, numerics in particular in 1D implies
that the real exponents are different from the one-loop results.
This has recently been explained in terms of two-loop corrections,
but the traditional interpretation has been in terms of 
``anomalous scaling'' \cite{Lesch,Juanma}, meaning 
that as $t \to \infty$, the typical
height difference between neighboring sites increases without limit. 


\vspace{1cm}
{\centerline{\bf ACKNOWLEDGMENTS}}
We acknowledge partial support
from the European Network contract ERBFMRXCT980183, from
the Academy of Finland's Center of Excellence Program,
and DGESIC (Spain) project PB97-0842. We thank R. Dickman, 
A. Vespignani, R. Pastor-Satorras, K. B. Lauritsen,
and S. Zapperi for useful discussions and long standing 
pleasant collaborations.

\end{multicols}


\begin{thebibliography}{99}       

\bibitem{Marro}
J. Marro and R. Dickman 
{\em Nonequilibrium Phase Transitions in Lattice Models} 
(Cambridge University Press, Cambridge, 1999).

\bibitem{Granada} See, G. Grinstein and
 M. A. Mu\~noz, {\it The Statistical
Mechanics of Systems with Absorbing States
}, in ``Fourth Granada Lectures in Computational Physics'',
edited by P. Garrido and J. Marro,
Lecture Notes in Physics, Vol. 493 (Springer, Berlin 1997),
p. 223, and references therein.            

\bibitem{Haye} H. Hinrichsen, Adv. Phys. {\bf 49}, 1, (2000).

\bibitem{DK} E. Domani and W. Kinzel, Phys. Rev. Lett. 
{\bf 53}, 311 (1984). 

   
\bibitem{FES}
A. Vespignani, R. Dickman, M. A. Mu\~noz, 
and S. Zapperi, 
Phys. Rev. Lett. {\bf 81}, 5676 (1998);
Phys. Rev. E {\bf 62}, 4564 (2000);
R. Dickman, M. A. Mu\~noz, A. Vespignani, and S. Zapperi,
Braz. J. Phys. {\bf 30}, 27 (2000).

\bibitem{HHZ}
T. Halpin-Healy and Y.-C. Zhang,
Phys. Rep. {\bf 254}, 215 (1995).

\bibitem{Barabasi}
A. -L. Barab\'asi and H. E. Stanley,
{\it Fractal Concepts in Surface Growth},
(Cambridge University Press, Cambridge, 1995).


\bibitem{RFT}
H. K. Janssen,
Z. Phys. {\bf 42}, 141 (1981); {\it ibid.} {\bf 58}, 311 (1985);
P. Grassberger, Z. Phys. B {\bf 47}, 465 (1982);
Cardy J.L., Sugar R.L. (1980), J. Phys. A {\bf 13}, L423.


\bibitem{BAW}
P. Grassberger, F. Krause, and T. von der Twer, 
J. Phys. A {\bf 17}, L105 (1984);
P. Grassberger,
{\it ibid}. {\bf 22}, L1103 (1989);
H. Takayasu and A. Yu. Tretyakov, 
Phys. Rev. Lett. {\bf 68}, 3060 (1992);
I. Jensen, 
Phys. Rev. E {\bf 50}, 3623 (1994);
N. Menyhard and G. \'Odor,
J. Phys. A {\bf 29}, 7739 (1996);
J. Cardy and U. C. T\"auber,
Phys. Rev. Lett. {\bf 77}, 4780 (1996).
 
 
\bibitem{dyp}   
J.L. Cardy, J. Phys. A {\bf 16}, L709 (1983);
J.L. Cardy and P Grassberger, J. Phys. A {\bf 18}, L267 (1985).          

\bibitem{WOH} 
F. van Wijland, K. Oerding, and H. J. Hilhorst, Physica A {\bf 251},
179 (1998). See also R. Kree, B. Schaub, and B. Schmittmann,
Phys. Rev. A {\bf 39}, 2214 (1989).


\bibitem{Dif} If the conserved field is diffusive, there is not an
infinite number of absorbing states, and the phenomenology
is completely different. See also:
J. E. de Freitas, L. S. Lucena, L. R. da Silva, and
H. J. Hilhorst, Phys. Rev. E {\bf 61}, 6330 (2000);
 J.-P. Leroy,  H. J. Hilhorst,  K. Oerding, and 
F. van Wijland, J. Stat. Phys. {\bf 99}, 1365 (2000),
and \cite{WOH}.


\bibitem{Romu1} M. Rossi, R. Pastor-Satorras, and A. Vespignani,
Phys. Rev. Lett. {\bf 85}, 1803 (2000).

\bibitem{Nattermann}
T. Nattermann, S. Stepanow, L.-H. Tang, and  H. Leschhorn,
J. Phys. (France) II {\bf 2}, 1483 (1992); H. Leschhorn,
T. Nattermann, S. Stepanow, and L.-H. Tang,
 Ann. Physik {\bf 7}, 1 (1997).

\bibitem{NF}
O. Narayan and D. S. Fisher,
Phys. Rev. B {\bf 48}, 7030 (1993).

\bibitem{Doussal} P. Chauve, P. Le Doussal, and K. J. Wiese,
Phys. Rev. Lett. {\bf 86}, 1785 (2001).  

\bibitem{Lesch}
H. Leschhorn, Physica A {\bf 195}, 324 (1993).

\bibitem{LT} H. Leschhorn and L.-H. Tang, Phys. Rev. Lett.
{\bf 70}, 2973 (1993).

\bibitem{TKD} L.-H. Tang, M. Kardar, and D. Dhar,
 Phys. Rev. Lett. {\bf 74}, 920 (1995).

\bibitem{columnar} G. Parisi and L. Pietronero,  Europhys. Lett. {\bf 16},
321 (1991); Physica A {\bf 179}, 16 (1991).
 
 \bibitem{BTW}
        P. Bak, C. Tang, and K. Wiesenfeld,
        Phys. Rev. Lett. {\bf 59}, 381 (1987);
        Phys. Rev. A {\bf 38}, 364 (1988).

\bibitem{Manna}
S. S. Manna, J. Phys. A {\bf24}, L363 (1992);
J. Stat. Phys. {\bf 59}, 509 (1990).
          
\bibitem{SOC} H. J. Jensen, {\it Self organized criticality},
(Cambridge Univ. Press, Cambridge, 1998).       
 G. Grinstein, in
{\it Scale Invariance, Interfaces and Nonequilibrium Dynamics},
{\it NATO Advanced Study Institute, Series B: Physics},
vol. 344, A. McKane et al., Eds.
(Plenum, New York, 1995).          

\bibitem{Romu2}  R. Pastor-Satorras, and A. Vespignani, Phys. Rev. E
{\bf 62}, 5875 (2000); See also Cond-mat/0101358. 


\bibitem{AIP} M. A. Mu\~noz, R. Dickman, R. Pastor-Satorras,
A. Vespignani, and S. Zapperi, ``Sandpiles and absorbing state
phase transitions: recent results and open problems,"
to appear in {\it Proceedings of the $6th$ Granada seminar on
computational physics}, Ed. J. Marro and P. L. Garrido. (American
Institute of Physics, 2001). e-print: cond-mat/0011447.

\bibitem{previous} 
O. Narayan and A. A. Middleton,
Phys. Rev. B {\bf 49}, 244 (1994);
M. Paczuski, S. Maslov, and P. Bak, 
Phys. Rev. E. {\bf 53}, 414 (1996).
M. Paczuski and S. Boettcher,
Phys. Rev. Lett. {\bf 77}, 111 (1996);     
L.~A.~N. Amaral and K.~B. Lauritsen,
Phys. Rev. E {\bf 54}, R4512 (1996); 
{\it ibid.\/} {\bf 56}, 231 (1997);
C. Tang and P. Bak,
Phys. Rev. Lett. {\bf 60}, 2347 (1988);
D. Cule and T. Hwa,
Phys. Rev. B {\bf 57}, 8235 (1998).


\bibitem{mapping}
K.~B. Lauritsen and M.~J. Alava, 
preprint cond-mat/{\-}9903349;
M. Alava and K. B. Lauritsen, Europhys. Lett.
{\bf 53}, 569 (2001);
M.~J. Alava and K.~B. Lauritsen, in preparation.         

\bibitem{marsili} M. Marsili,
J. Stat. Phys. {\bf 77}, 733 (1994).

\bibitem{BS} M. Felici, G. Caldarelli, A. Gabrielli,
and L. Pietronero, Phys. Rev. Lett. {\bf 86}, 1896 (2001).

\bibitem{example} For example, in invasion
percolation the bond threshold is thus related to the time the
bond remains active on the boundary, and the quenched variable
is replaced with an annealed one, with a memory. In sandpiles,
the local dynamics can be mapped to quenched disorder via almost
exactly the inverse route.

                                                     
\bibitem{DM} R. Dickman and M. A. Mu\~noz, 
Phys. Rev. E {\bf 62}, 7632 (2000). 

\bibitem{Fock} 
M. Doi, J. Phys. A {\bf 9}, 1465 (1976); 
L. Peliti, J. Phys. I {\bf 46}, 1469 (1985);
B. P. Lee and J. L. Cardy J. Stat. Phys. {\bf 80}, 971 (1995).

\bibitem{RG} D. J. Amit, {\em Field Theory, the Renormalization Group and Critical Phenomena}, 
(World Scientific, Singapore, 1992). J. Zinn- Justin, {\it Quantum field theory
and critical phenomena}, Clarendon Press, (Oxford), 1994.        
            

\bibitem{1d} We believe that effects like the ones described here
are at the basis of the discrepancy between the one-dimensional 
Manna model and the 1-d LIM class. R. Dickman, M. Alava, M. A. Mu\~noz,
J. Peltola, A. Vespignani, and S. Zapperi, p
reprint 2001 (cond-mat/0101381).   


\bibitem{FV}
F. Family and T. Vicsek, J. Phys. A  {\bf 18}, L75 (1985).
           

\bibitem{DPD} L.-H- Tang and H. Leschhorn, Phys. Rev. A {\bf 45}, R8309 
(1992);  S. Buldyrev et al. Phys. Rev. A {\bf 45}, R8313 (1992);
Z. Olami, I. Procaccia, and R. Zeitak, Phys. Rev. E {\bf 49}, 1232 (1994).

\bibitem{BGM}  A. L. Barabasi, G. Grinstein, and M. A. Mu{\~n}oz,
Phys. Rev. Lett. {\bf 76}, 1481, (1996).        

                                                     
\bibitem{DM} R. Dickman and M. A. Mu\~noz, 
Phys. Rev. E {\bf 62}, 7632 (2000). 


\bibitem{Juanma} 
J. M. L\'opez, Phys. Rev. Lett. {\bf 83}, 4594 (1999);
J. M. L\'opez and  M. A. Rodr{\'\i}guez,
Phys. Rev. E {\bf 54}, R2189 (1996);
J. M. L\'opez, M. A. Rodr{\'\i}guez and R. Cuerno,
Phys. Rev. E {\bf 56}, 3993 (1997).
                                     
  
\end{thebibliography}
\end{document}